\documentclass[10pt,letterpaper,twocolumn]{article} %% two column, final layout

\usepackage{ol2}
\usepackage[draft]{hyperref}
\usepackage{amsmath}

\begin{document}

\twocolumn[ %% activate for two-column option

\title{Anderson localization without eigenstates in photonic quantum walks}

%% For REVTeX it is possible to automate superscript and e-mail callouts with the superscriptaddress option; see REVTeX4 documentation.

\author{Stefano Longhi}
\address{Dipartimento di Fisica, Politecnico di Milano and Istituto di Fotonica e Nanotecnologie del Consiglio Nazionale delle Ricerche, Piazza L. da Vinci 32, I-20133 Milano, Italy (stefano.longhi@polimi.it)}
\address{IFISC (UIB-CSIC), Instituto de Fisica Interdisciplinar y Sistemas Complejos, E-07122 Palma de Mallorca, Spain}

\begin{abstract}
Anderson localization is ubiquitous in wavy systems with strong static and uncorrelated disorder. The delicate destructive interference underlying Anderson localization is usually washed out in the presence of temporal fluctuations or aperiodic drives in the Hamiltonian, leading to delocalization and restoring transport. However, in one-dimensional lattices with off-diagonal disorder Anderson localization can persist for arbitrary time-dependent drivings that do not break a hidden conservation law originating from the chiral symmetry, leading to the dubbed "localization without eigenstates". Here it is shown that such an intriguing phenomenon can be observed in discrete-time photonic quantum walks with static disorder applied to the coin operator, and can be extended to non-Hermitian dynamics as well.
\end{abstract}

%\ocis{130.2790,  130.3120, 000.160)}
 ] %% activate for two-column option

Anderson localization \cite{r1} is a ubiquitous wave phenomenon that arises due to a delicate destructive interference of waves scattered off by uncorrelated static disorder. Since its discovery, it has been instrumental
for the understanding of a plethora of physical phenomena \cite{r2,r3,r4}, with experimental demonstrations reported in many systems, including photonics \cite{r5,r6,r7,r8,r9,r10,r12,r13,r14,r15,r16,r17,r18,r19,r20,r21}. A common belief is that dephasing effects arising from the interaction with an environment, as well as in time-varying disordered systems with broken time translation symmetry, Anderson localization is destroyed and transport is restored \cite{r23,r24,r25b,r25}: only in periodically or quasi-periodically driven systems, where the time behaves as an additional synthetic dimension, Floquet-Anderson localization can be observed \cite{r26,r27,r28,r29}. However,  in disordered lattices with chiral symmetry one can construct drives with a hidden conservation law such that Anderson localization persists indefinitely for {\em arbitrary} aperiodic
drives \cite{r30}. Such a kind of persistent localization with aperiodic drives, dubbed "localization without eigenstates" \cite{r30} because of the absence of truly eigenstates like in static or periodically-driven (Floquet) systems, has remained elusive to experimental observations so far.\\
In this Letter we extend the idea of "localization without eigenstates" to disordered discrete-time systems, and suggest a photonic quantum walk (QW) setup, with static spatial disorder in the coin operators and preserved chiral symmetry, as an experimentally feasible platform for the observation of such an intriguing kind of persistent localization.\\
To introduce the main idea of "localization without eigenstates" and the role of chiral symmetry, let us first consider the single-particle one-dimensional Anderson model with off-diagonal disorder perturbed by a time dependent term with arbitrary time-dependence $f(t)$, describe by the Hamiltonian \cite{r30}
\begin{equation}
\hat{H}(t)= \sum_n J_{n} \left( | n \rangle \langle n+1|+ | n+1 \rangle \langle n| \right)+ f(t) \hat{P}  \equiv \hat{H}_0 + f(t) \hat{P} 
\end{equation} 
where $J_n$ are the hopping amplitudes, which are assumed to be uncorrelated random variables with some probability distribution function, and $\hat{P}$ is a time-independent local operator defined by $\hat{P}=\sum_{n,m} P_{n,m} |n \rangle \langle m|$ in the Wannier basis $|n \rangle$. For a given initially-localized state $| \psi(t=0) \rangle= | \psi_0 \rangle$, the state vector of the system at time $t$ reads $| \psi(t) \rangle=\hat{U}(t) | \psi_0 \rangle$, where the propagator is given by the time-ordered integral $\hat{U}(t)= \mathcal{T} \exp [-i\int_0^t dt' \hat{H}(t') ]$. The spreading of excitation in the lattice can be measured by the second moment \cite{r24} $\sigma^2(t)=\sum_n n^2 | \langle n | \psi(t) \rangle|^2$. Localization "without eigenstates" corresponds to $\sigma^2(t) \leq M$, with a finite $M$, for any time $t$ and arbitrary initial excitation of the lattice.  When $f(t)=0$, i.e. $\hat{H}(t)=\hat{H}_0$, the static off-diagonal disorder introduces Anderson localization, albeit the eigenstates with energy $E=0$ show a sub-exponential localization with a  diverging localization length \cite{r31,r32,r33}. The system displays chiral (sublattice) symmetry, namely $ \{ \hat{H}_0, \hat{L} \}=\hat{H}_0 \hat{L}+\hat{L}\hat{H}_0 =0$, where $\hat{L}=\sum_n (-1)^n |n \rangle \langle n |$ and $\hat{L}^2=\mathcal{I}$. This means that, if $| \epsilon_k \rangle$ is an eigenstate of $\hat{H}_0$ with eigenenergy $\epsilon_k$, then $\hat{L}| \epsilon_k \rangle \equiv | \tilde{\epsilon}_k \rangle$ is an eigenstate of $\hat{H}_0$ with eigenenergy $-\epsilon_k$. $\hat{H}_0$ can be thus diagonalized as $\hat{H}_0=\sum_k \epsilon_k (| \epsilon_k \rangle \langle \epsilon_k|-| \tilde{\epsilon}_k \rangle \langle \tilde{\epsilon}_k|)$. The addition of an arbitrary local perturbation $\hat{P}$ with aperiodic driving amplitude $f(t)$ results rather generally in delocalization. However, as shown in \cite{r30}  for the special local perturbation $\hat{P}=\hat{L}$, i.e. for $P_{n,m}=(-1)^n \delta_{n,m}$, localization persists for an arbitrary time dependence of the driving amplitude $f(t)$ as a result of the conservation law $ \{ \hat{H_0},\hat{P} \}=0$. In fact, taking into account that one can write $\hat{L}=\sum_k ( | \epsilon_k \rangle \langle \tilde{\epsilon}_k | + | \tilde{\epsilon}_k \rangle \langle {\epsilon}_k | )$
the chiral disordered model $\hat{H}(t)=\hat{H}_0+f(t) \hat{L}$ reduces to a collection of decoupled two-level systems, $| \epsilon_k \rangle$ and $| \tilde{\epsilon}_k \rangle$, where the local perturbation $f(t) \hat{L}$ introduces a coupling between the paired states  $| \epsilon_k \rangle$ and $| \tilde{\epsilon}_k \rangle$ with the same $k$. Consequently, the coupling between the paired states 
will not significantly change the localization properties of the time evolution operator $\hat{U}(t)$ at any time $t$ for arbitrary forms of the driving profile $f(t)$, resulting in the dubbed "localization without eigenstates" \cite{r30}.{\color{black} We note that addition of on-site potential disorder in the Hamiltonian (1) would break chiral symmetry and would mix different $k$ subspaces, resulting rather generally in delocalization of eigenstates of the time evolution operator \cite{r30}. Photonic implementations of the Anderson model (1) is possible in different setups, including coupled waveguide arrays and fibers (see e.g. \cite{  r7,r10,r12,r13,Referee1} ).  }\par
We can extend the previous idea to discrete-time QWs \cite{r33b}, which 
can be realized using different platforms, such as
polarizer beam splitters and quarter-wave plates \cite{r9,r15,r20,r21,r34,r35}, fiber
network loops \cite{
 r17,r19,r21,r36,r37,r37b}, or chiral light carrying orbital angular
momentum \cite{r38,r39,r40}.
{\color{black} As compared to continuous-time QWs in coupled waveguide lattices, discrete-time QWs offer the advantage of a simpler implementation of non-Hermitian dynamics \cite{r19,r20} and the ability of monitoring wave spreading at long propagation times (up to few thousands of time steps \cite{r19})}. In a QW the state vector is defined by 
 \begin{equation}
 | \psi(t) \rangle= \sum_n \left( u_n(t) |n \rangle \otimes |H \rangle+v_n(t) |n \rangle \otimes |V \rangle \right),
 \end{equation}
 where $n$ is the spatial position of the walker on a one-dimensional lattice and $H,V$ denote the internal degree of freedom of the walker (for example the horizontal H or vertical V polarization state of the photon). The state vector 
 evolves according to $| \psi (t+1) \rangle= \hat{\Theta}(t) | \psi (t) \rangle$, where the one-step propagator $\hat{\Theta}(t)$ is given by the composition of  three main operation: the conditional spatial shift operator $\hat{S}$, the spatial-dependent coin operator $\hat{C}$, and the phase shift operator $\hat{P}$. {\color{black} Disorder can be rather generally either stochastic or deterministic (quasiperiodic), and can be introduced either in space or time \cite{Referee2}.  }
  %The latter two operators act on the internal degree of freedom (coin state) of the walker, while the former one acts on the spatial position degree of the walker.  
   In our model, we introduce static spatial disorder in the  coin operator $\hat{C}$ \cite{r20}, while the time-dependence of the propagator arises from the $t$-dependence of $\hat{P}$. Namely, let us assume $\hat{\Theta}(t)=\hat{P}(t) \hat{S} \hat{C}$ with
 \begin{equation}
 \hat{S}  =  \sum_n \left( |n-1 \rangle \langle n | \otimes |H \rangle \langle H |+ |n+1 \rangle \langle n | \otimes |V \rangle \langle V | \right),
 \end{equation}
 \begin{equation}
 \hat{C}  =  \sum_n  \left(
 \begin{array}{cc}
\cos \theta_n &  \sin \theta_n \\ 
 - \sin \theta_n & \cos \theta_n 
 \end{array} \right)  \otimes | n \rangle \langle n |
 \end{equation}
 \begin{eqnarray}
 \hat{P}(t)  & = &  [1-f(t)] 
\sum_n \left(
  \begin{array}{cc}
1 & 0 \\ 
0 & 1 
 \end{array} \right) \otimes | n \rangle \langle n | + \\
 & + &f(t) \sum_n 
  \left(
  \begin{array}{cc}
\exp(i n \varphi) & 0 \\ 
0 & \exp(-i \varphi n) 
 \end{array} \right) \otimes |n \rangle \langle n| . \nonumber
 \end{eqnarray}
In the above equations, $\theta_n$ are the space-dependent rotation angles of coin state, $\varphi$ a phase gradient term, and $f(t)$ is a function of discrete time $t$ that can take only the two values $f(t)=0,1$. Note that when $f(t)=0$ the operator $\hat{P}(t)$ reduces to the identity operator, while when $f(t)=1$ a gradient phase $\pm n \varphi$ is applied to the internal states H and V. Since the sequence $f(t)$ can be rather generally aperiodic in discrete time $t$ and even stochastic, the system does not possess eigenstates. 
The discrete-time evolution of the system, from an initially-localized state $| \psi(t=0) \rangle = | \psi_0 \rangle$, reads $| \psi(t) \rangle = \hat{U}(t) | \psi_0 \rangle$, where the evolution operator is given $\hat{U}(t)= \hat{\Theta}(t-1) \times  \hat{\Theta}(t-2) \times ... \hat{\Theta}(1)$. The spreading of excitation in the lattice can be measured by the second moment $\sigma^2(t)=\sum_n n^2 \{ |u_n(t)|^2+|v_n(t)|^2\}$, and localization corresponds to $\sigma^2(t) \leq M$, with a finite $M$, for any time $t$ and arbitrary initial excitation.  The localization properties of the QW largely depend on the choice of the phase $\varphi$. We basically have three distinct dynamical regimes.\\
(i) {\em The ordinary Anderson localization regime.} For $f(t) \equiv0$, the propagator $\hat{\Theta}=\hat{S} \hat{C}$ is time-independent and, for a rather arbitrary form of uncorrelated disorder  of the coin angles $\theta_n$, one observes Anderson localization, with all the eigenstates of $\Theta$ exponentially localized with a finite localization length \cite{r41}; {\color{black}  other types of disorder, such as deterministic
aperiodic sequences, are not considered here since they would not lead to strong localization \cite{Referee2}.}  Additionally, $ \{ \hat{\Theta}, \hat{L} \}=\hat{\Theta} \hat{L}+\hat{L}\hat{\Theta} =0$, where $\hat{L}=\sum_n (-1)^n |n \rangle \langle n | \otimes (|H\rangle \langle H | +|V \rangle \langle V |)$ and $\hat{L}^2=\mathcal{I}$. This implies that, if $| \epsilon_k \rangle$ is a localized eigenstate of $\hat{\Theta}$ with quasi energy $\mu_k$ ($ -\pi \leq \mu_k < \pi$), i.e. $\hat{\Theta} | \epsilon_k \rangle= \exp( -i \mu_k)  | \epsilon_k \rangle$, then $\hat{L}| \epsilon_k \rangle \equiv | \tilde{\epsilon}_k \rangle$ is an eigenstate of $\hat{\Theta}$ with quasi energy $\mu_k+ \pi$. {\color{black}This kind of symmetry plays an analogous role than the single-particle chiral symmetry in the continuous-time QW introduced for Eq.(1) \cite{r30}.} 
 \begin{figure}
  \centering
    \includegraphics[width=0.48\textwidth]{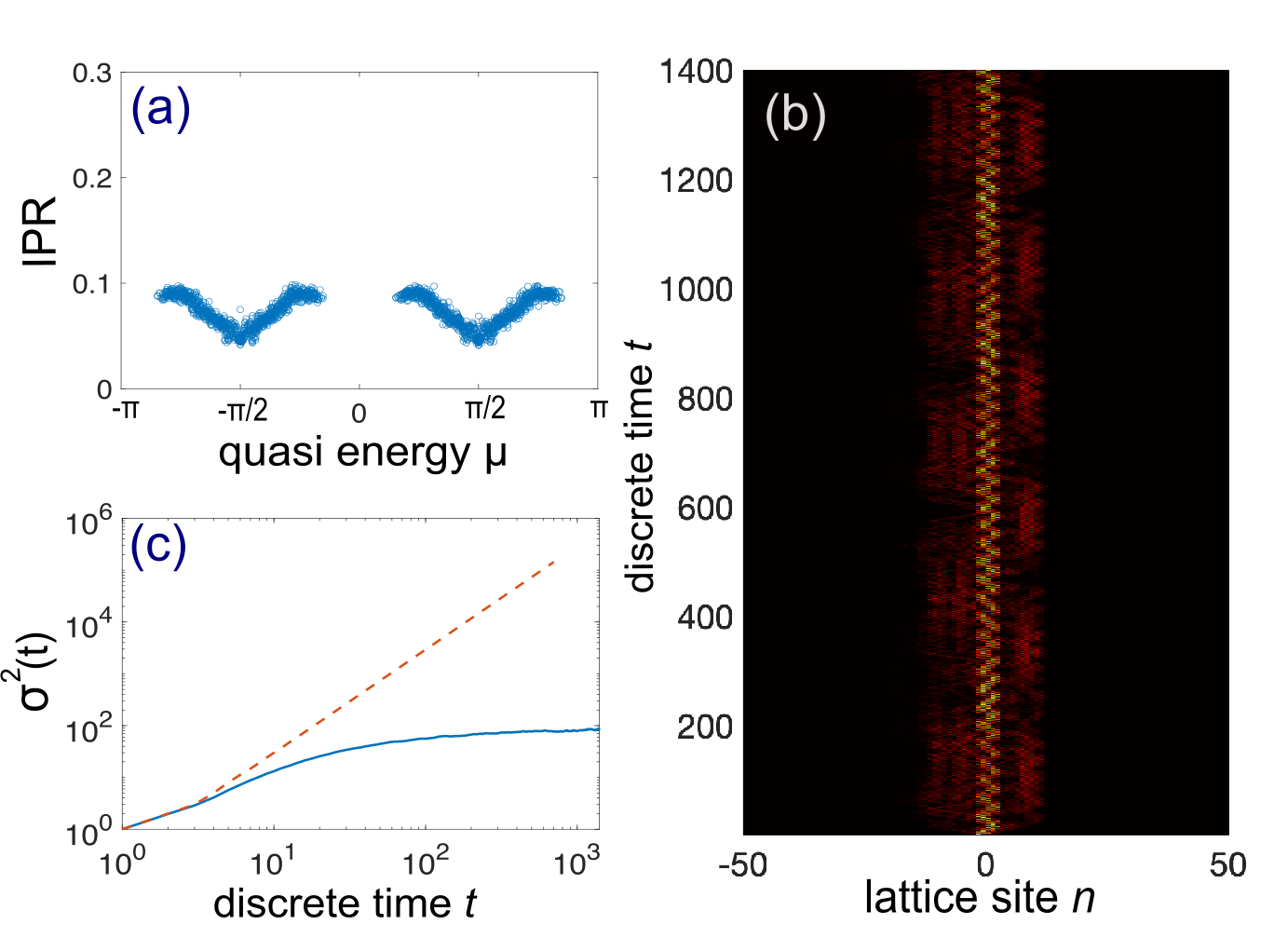}
    \caption{Anderson localization in the discrete-time QW with static spatial disorder in the coin operators, respecting chiral symmetry, and $f(t)=0$. (a) Numerically-computed IPR versus quasi energy $\mu$ of the eigenstates in a lattice comprising $L=1000$ sites. The coin angle $\theta_n$ is assumed to be uniformly distributed in the range $(\pi/16, 7 \pi / 16)$. The IPR is averaged over 200 different realizations of disorder. (b) Typical example of temporal evolution of the QW on a lattice for a given realization of disorder. At initial time the walker is localized at site $n=0$ with internal state $H$. (c) Numerically-computed  behavior of the second moment $ \overline{\sigma^2(t)}$ versus discrete time $t$ (solid curve); statistical average is made over 200 realizations of disorder. The red dashed curve shows, for comparison, the behavior of $\sigma^2(t)$ in a disorder-free QW with $\theta_n= \pi/4$, displaying ballistic transport $\sigma^2(t) \sim t^2$.}
\end{figure} 
The Anderson localization in this regime is clearly illustrated in Fig.1.
%where the coin angles $\theta_n$ are assumed to be uncorrelated random variables with uniform distribution in the range $(\tilde{\theta}_1, \tilde{\theta}_2)$, with $\tilde{\theta}_1=\pi /16$ and $\tilde{\theta}_2=7 \pi/16$ . 
The localization properties of the eigenstates $ |\epsilon_k \rangle$ are measured by the inverse participation ratio (IPR) \cite{r24}. Assuming $\langle \epsilon_k | \epsilon_k \rangle=1$, the IPR reads
\begin{equation}
IPR_k= \sum_n  | \langle H| \otimes \langle n | \epsilon_k \rangle|^4 +| \langle V| \otimes \langle n | \epsilon_k \rangle|^4.
%IPR_k= \frac{\sum_n  | \langle H| \otimes \langle n | \epsilon_k \rangle|^4 +| \langle V| \otimes \langle n | \epsilon_k \rangle|^4}{  \left( \sum_n | \langle H| \otimes \langle n | \epsilon_k \rangle|^2 +| \langle V| \otimes \langle n | \epsilon_k \rangle|^2 \right)^2}.
\end{equation}
In a lattice of large size $L$, the IPR vanishes as  $\sim 1/L$ for an extended state, whereas it remains finite (of order $ \sim 1$) for a localized state. 
Figure 1(a) shows the numerically-computed behavior of the IPR versus the quasi energy $\mu$ of the  eigenstates in a lattice of size $L=1000$, averaged over 200  different realizations of disorder. Note that for all the eigenstates the IPR remains well above zero, indicating the spectral localization of $\hat{\Theta}$. 
Figure 1(b) depicts a typical QW spreading dynamics on a pseudocolor map,  for a given realization of disorder and for the initial condition $ | \psi_0 \rangle=  \sum_n \delta_{n,0} |n \rangle  \otimes | H \rangle$, corresponding to the walker at site $n=0$ with the internal state $H$. Figure1(c) shows the behavior of the second moment $ \overline{ \sigma^2(t)}$ versus discrete time on a log scale, where the overbar denotes a statistical average over 200 realizations of disorder. The dashed curve in Fig.1(c) shows, for comparison, the corresponding behavior of $\sigma^2(t)$ in the absence of disorder and for $\theta_n= \pi/4$ (the Hadamard coin), displaying ballistic transport $\sigma^2(t) \sim t^{\delta}$ with an exponent $\delta=2$. Clearly, in the presence of disorder dynamical localization is observed.\\
 \begin{figure}
  \centering
    \includegraphics[width=0.48\textwidth]{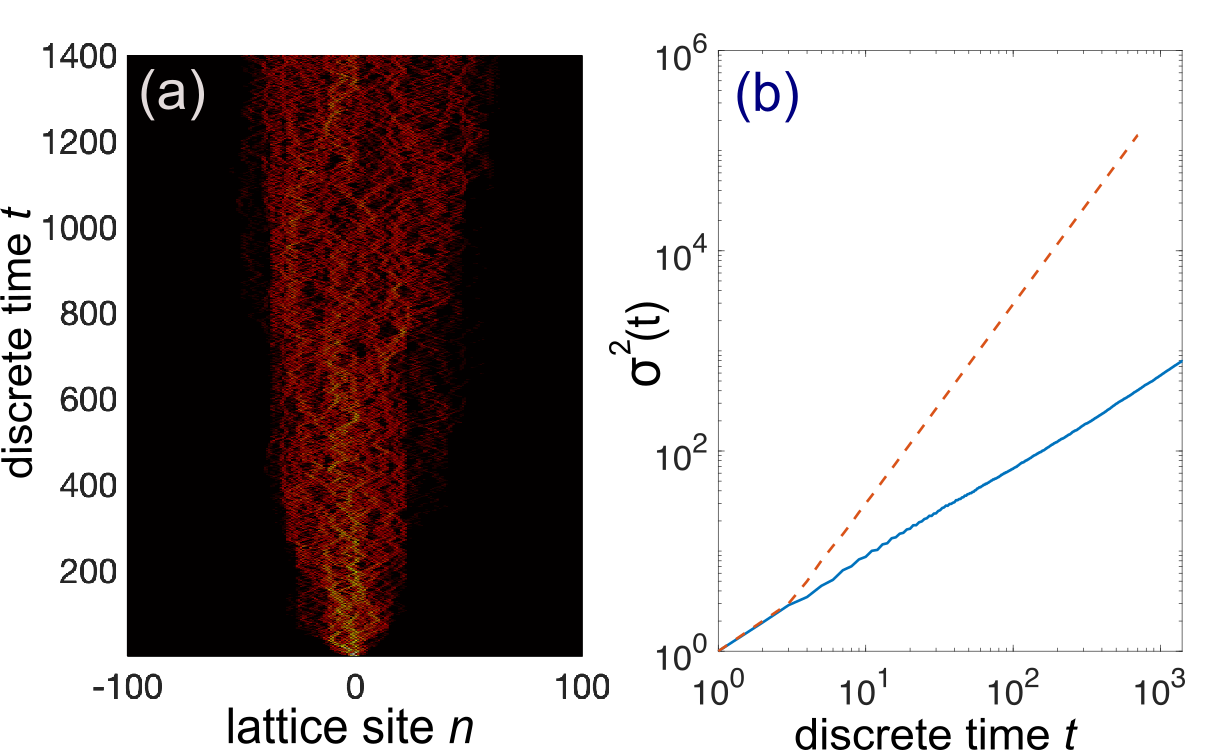}
    \caption{Diffusive transport induced by a time-varying phase shift operator for $\varphi= \pi/2$. At each time step, $f(t)$ can take the two values $0$ or $1$ with equal probabilities. Other conditions, i.e. static disorder, lattice size etc.  are as in Fig.1.  (a) Typical example of temporal evolution of the QW on the lattice for a given realization of disorder. At initial time the walker is localized at site $n=0$ with internal state $H$. (b) Numerically-computed  behavior of the second moment $ \overline{\sigma^2(t)}$ versus discrete time $t$ (solid curve); statistical average is made over 200 realizations of static disorder. 
    Note that $\overline{\sigma^2(t)} \sim t^{\delta}$ with exponent $\delta \simeq 1$, indicating diffusive transport. The red dashed curve shows, for comparison, the behavior of $\sigma^2(t)$ in a disorder-free lattice with $\theta_n= \pi/4$ and $f=0$.}
\end{figure} 
 \begin{figure}
  \centering
    \includegraphics[width=0.48\textwidth]{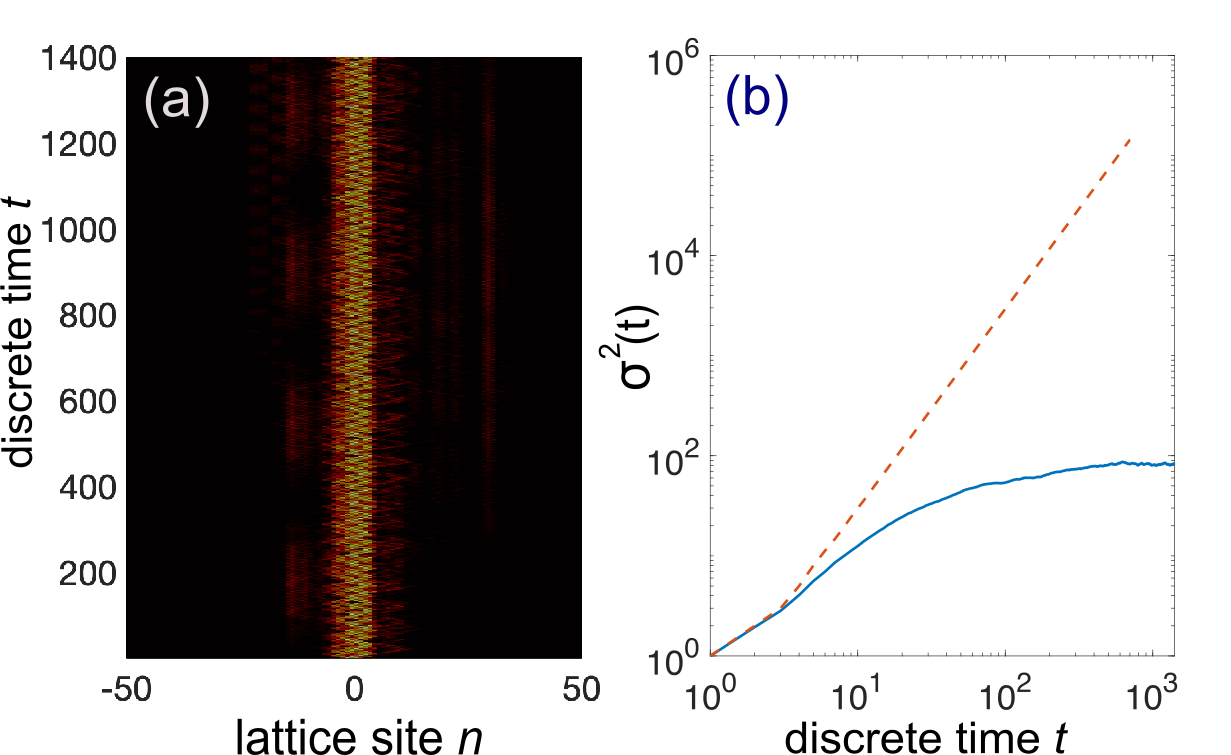}
    \caption{Persistent localization without eigenstates. Same as Fig.2, but for a phase gradient $\varphi= \pi$. Note that Anderson localization persists in this regime.}
\end{figure} 
 (ii) {\em Diffusive transport induced by the time-varying phase shift operator.} Let us now assume that the dynamics is non-autonomous and the function $f(t)$ is highly aperiodic to break discrete time translational symmetry. For the sake of definiteness, let us assume for example that $f(t)$ is stochastic and can take, at each time step $t$, either the value 0 or 1 with the same probability. In other words, $f(t)$ are independent stochastic variables with a Bernoulli distribution. For a rather arbitrary value of the phase gradient $\varphi$, namely for $\varphi$ far from $0, \pi$, numerical simulations indicate that the Anderson localization observed in the static case is washed out and diffusive-like transport is restored, the second moment growing in time like $\sigma^2(t) \sim t^ \delta$ with an exponent $\delta \sim 1$, characteristic of diffusive transport. This behavior is illustrated in Fig.2 for $\varphi= \pi/2$, however similar behavior is found for a wide interval of $\varphi$ around $\pi/2$, except for $\varphi$ approaching the boundaries $0$ and $\pi$. The diffusive-like transport observed in the numerical simulations is typical  of time-dependent Anderson Hamiltonians \cite{r24} and arises from the coupling among localized Anderson states $| \epsilon_k \rangle$, $ | \tilde{\epsilon}_k \rangle$ with different $k$ indices induced by the time-dependent shift operator $\hat{P}(t)$, which breaks the chiral symmetry of the undriven system when $\varphi \neq 0, \pi$.\\
(iii) {\em Persistent localization without eigenstates}. When the phase gradient $\varphi$ entering in the phase shift operator $\hat{P}(t)$ is tuned to the value $ \pi$, one has $\hat{P}(t)= \mathcal{I}$ (the identity operator) when $f(t)=0$, or $\hat{P}(t)=\hat{L}$ when $f(t)=1$. Therefore, for such a special value of $\varphi$ at each propagation step the evolution operator $\hat{\Theta}(t)$, which is either $\hat{S} \hat{C}$ or $\hat{L} \hat{S} \hat{C}$, anti-commutes with $\hat{L}$, and  --akin to the continuous-time model discussed in the introductory section-- the discrete-time evolution of the system is described by  a set of decoupled two-level systems, with paired states $| \epsilon_k \rangle$ and $| \tilde{\epsilon}_k \rangle= \hat{L} | \epsilon_k \rangle$ involving the same index $k$. More precisely, if we expand the state vector of the system as a superposition of the localized eigenstates $| \epsilon_k \rangle$ and $| \tilde{\epsilon}_k \rangle$ of $\hat{S}\hat{C}$, i.e. after letting
\begin{equation}
| \psi (t) \rangle = \sum_k \left\{ a_k(t) | \epsilon_k \rangle - b_k(t) | \tilde{\epsilon}_k \rangle\right\} \exp(-i \mu_k t) ,
\end{equation}
the discrete-time evolution of the two-level amplitudes $a_k(t)$ and $b_k(t)$ for the paired states read
\begin{eqnarray}
a_k(t+1) & = & [1-f(t)]a_k(t)-f(t) b_k(t) \\
b_k(t+1) & = & [1-f(t)]b_k(t)-f(t) a_k(t) .
\end{eqnarray}
Clearly, the coupling between the paired states with same $k$ will not significantly change the localization properties of the time evolution operator $\hat{U}(t)$ at any discrete time $t$, resulting in persistent localization without eigenstates. This result is illustrated in Fig.3. {\color{black}  The number of time steps requited to observe localization, for parameter values chosen in the simulations, is $ \sim 500$; a shorter number of time steps could be obtained, if needed, by working with a mean rotation angle $\theta$ closer to $ \pi/2$.}\\
Finally, let us observe that the constraint assumed for the allowed values of $f(t)$, i.e. $f(t)=0$ or $f(t)=1$, comes from the need to keep the phase shift operator $\hat{P}(t)$ unitary, which is the case of an Hermitian photonic QW. However, extending the analysis to non-Hermitian QWs \cite{r19,r20,r37} and considering gain and/or loss terms at spatial sites $|n \rangle$, the discrete-time function $f(t)$ at each step can take rather arbitrary values. In this case for $\varphi=\pi$ one has
$\hat{P}(t) =  \left( \sum_n |  n \rangle \langle  n |- 2 f(t) \sum_n |2n+1 \rangle \langle 2n+1| \right) \otimes  (|H \rangle \langle H|+|V \rangle \langle V|)$.
The operator $\hat{P}(t)$ basically corresponds to the application, at odd sites of the lattice, of a loss/gain modulation amplitude $\gamma(t)$ such that $\exp[-\gamma(t)]=1-2f(t)$. 
As an example, Fig.4 shows persistent localization as obtained by applying at odd sites of the lattice the loss/gain modulation $\gamma(t)=A \cos (2 \pi \alpha t)$, with $\alpha$ irrational.\\  
%Therefore, photonic QWs could provide an accessible platform for the observation of persistent localization without eigenstates in the non-Hermitian realm as well.\\
 \begin{figure}
  \centering
    \includegraphics[width=0.48\textwidth]{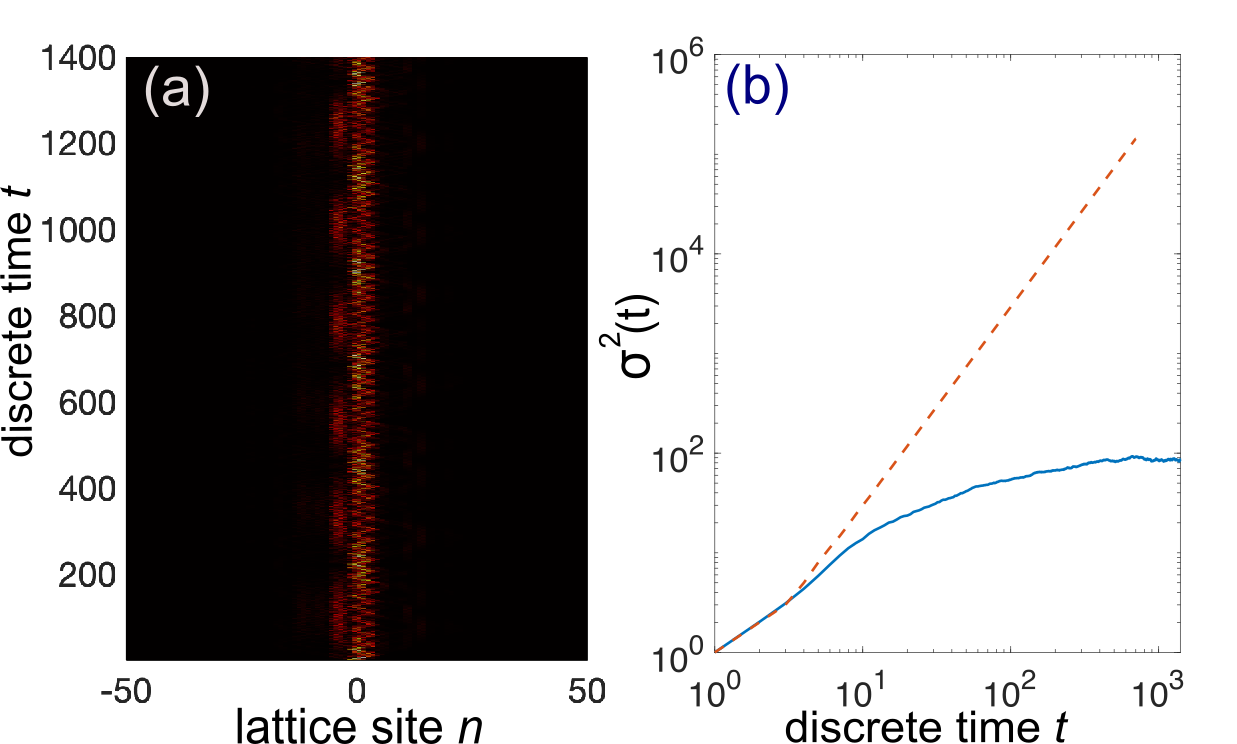}
    \caption{Persistent localization in a non-Hermitian QW. At each time step, a loss/gain amplitude $\exp[-\gamma(t)]$ is impressed to the odd sites of the lattice, with $\gamma(t)=A \cos ( 2 \pi \alpha t)$, $\alpha= (\sqrt{5}-1)/2$ and $A=0.2$. Since the non-Hermitian QW does not conserve the norm, to compute the second moment $\sigma^2(t)$ the state vector $| \psi(t) \rangle$ is renormalized at each time step.}
\end{figure}

To conclude, we predicted persistent Anderson localization without eigenvalues in discrete-time photonic QW with disorder in the coin operator. Our results suggest that photonic QWs could provide an accessible platform for the observation of persistent localization protected by chiral symmetry,  and that such an intriguing phenomenon could be extended to non-Hermitian dynamics as well.\\
\\
\noindent
{\bf Disclosures}. The author declares no conflicts of interest.\\
{\bf Data availability}. No data were generated or analyzed in the presented research.\\
{\bf Funding}. Agencia Estatal de Investigacion (MDM-2017-0711).
%\\
%{\bf Supplemental document}. See Supplement 1 for supporting content.


\begin{thebibliography}{99}


%%%%%%%%%%%%%%%%%%%%%%%%%%%%%%%
% References (short version)  %
%%%%%%%%%%%%%%%%%%%%%%%%%%%%%%%

 
\bibitem{r1}
P. W. Anderson, Phys. Rev. {\bf 109}, 1492 (1958).
\bibitem{r2}
S. Fishman, D.R. Grempel, and R.E. Prange, Phys. Rev. Lett. {\bf 49}, 509 (1982).
\bibitem{r3}
2. F. Evers and A.D. Mirlin, Rev. Mod. Phys. {\bf 80}, 1355 (2008).
\bibitem{r4}
A. Lagendijk, B. van Tiggelen, and D.S. Wiersma,  Phys. Today {\bf 62}, (8) 24 (2009).

\bibitem{r5}
D. S. Wiersma, P. Bartolini, A. Lagendijk, and R. Righini, Nature {\bf 390}, 671 (1997).
\bibitem{r6}
T. Schwartz, G. Bartal, S. Fishman, and M. Segev, Nature {\bf 446}, 52 (2007).
\bibitem{r7}
Y. Lahini, A. Avidan, F. Pozzi, M. Sorel, R. Morandotti, D.N. Christodoulides, and Y. Silberberg, Phys. Rev. Lett. {\bf 100}, 013906 (2008).
\bibitem{r8}
L. Sapienza, H. Thyrrestrup, S. Stobbe, P.D. Garcia, S. Smolka, and P. Lodahl, Science {\bf 327}, 1352 (2010).
\bibitem{r9}
A. Schreiber, K. N. Cassemiro, V. Potocek, A. Gabris, I. Jex, and Ch. Silberhorn, Phys. Rev. Lett. {\bf 106}, 180403 (2011).
\bibitem{r10}
L. Martin, G. Di Giuseppe, A. Perez-Leija, R. Keil, F. Dreisow, M. Heinrich, S. Nolte, A. Szameit, A.F. Abouraddy, D.N. Christodoulides, and B.E.A. Saleh, Opt. Express {\bf 19}, 13636 (2011).
%\bibitem{r11}
%D. Jovic, C. Denz, and M. Belic,  Opt. Photon. News {\bf 22}, 34 (2011).
\bibitem{r12}
U. Naether, Y.V. Kartashov, V.A. Vysloukh, S. Nolte, A. T\"unnermann, L. Torner, and A. Szameit, Opt. Lett. {\bf 37}, 593 (2012).
\bibitem{r13}
S. St\"utzer, Y. V. Kartashov, V. A. Vysloukh, A. T\"unnermann, S. Nolte, M. Lewenstein, L. Torner, and A. Szameit,  Opt. Lett. {\bf 37}, 1715 (2012).
\bibitem{r14}
M. Segev, Y. Silberberg, and D.N. Christodoulides, Nature Photon. {\bf 7}, 197 (2013).
\bibitem{r15}
A. Crespi, R. Osellame, R. Ramponi, V. Giovannetti, R. Fazio, L. Sansoni, F. De Nicola, F. Sciarrino, and P. Mataloni, Nature Photon. {\bf 7}, 322 (2013).
\bibitem{r16}
C. Cedzich, T. Rybar, A. H. Werner, A. Alberti, M. Genske, and R. F. Werner, Phys. Rev. Lett. {\bf 111}, 160601 (2013).
\bibitem{r17}
I.D. Vatnik, A. Tikan, G. Onishchukov, D.V. Churkin, and A.A. Sukhorukov, Sci. Rep. {\bf 7}, 4301 (2017).
\bibitem{r18}
M. Lee, J. Lee, S. Kim, S. Callard, C. Seassal, and H. Jeon,  Sci. Adv. {\bf 4}, e1602796 (2018).
\bibitem{r19}
S. Weidemann, M. Kremer, S. Longhi, and A. Szameit, Nature Photon. {\bf 15}, 576 (2021).
\bibitem{r20}
Q. Lin, T. Li, L. Xiao, K. Wang, W. Yi, and P. Xue,  Nature Commun. {\bf 13}, 3229 (2022).
\bibitem{r21}
A. Dikopoltsev, S. Weidemann, M. Kremer, A. Steinfurth, H.H. Sheinfux, A. Szameit, and M. Segev,
Sci. Adv. {\bf 8}, eabn7769 (2022).
%\bibitem{r22}
%J. Gao, Z.-S. Xu, D.A. Smirnova, D. Leykam, S. Gyger, W.-H. Zhou, S. Steinhauer, V. Zwiller, and A.W. Elshaari,  
%Phys. Rev. Res. {\bf 4}, 033222 (2022).\\
\bibitem{r23}
D.E. Logan and P.G. Wolynes, Phys. Rev. B {\bf 36}, 4135 (1987).
\bibitem{r24}
D.A. Evensky, R.T. Scalettar, and P.G. Wolynes,  J. Chem. Phys. {\bf 94}, 1149 (1990).
\bibitem{r25b}
L.Levi, Y. Krivolapov, S. Fishman, and M. Segev, Nature Phys. {\bf 8}, 912 (2012).
\bibitem{r25}
S. Gopalakrishnan, K.R. Islam, and M. Knap, Phys. Rev. Lett. {\bf 119}, 046601 (2017).
\bibitem{r26}
H. Hatami, C. Danieli, J. D. Bodyfelt, and S. Flach, Phys. Rev. E {\bf 93}, 062205 (2016).
\bibitem{r27}
R. Ducatez and F. Huveneers, Ann. Henri Poincar\'e {\bf 18}, 2415 (2017).
\bibitem{r28}
K. Agarwal, S. Ganeshan, and R.N. Bhatt, Phys. Rev. B {\bf 96}, 014201 (2017).
\bibitem{r29}
M.M. Wauters, A. Russomanno, R. Citro, G.E. Santoro, and L. Privitera, Phys. Rev. Lett. {\bf 123}, 266601 (2019).
\bibitem{r30}
H. Zhao, F. Mintert,  J. Knolle, and R. Moessner, Phys. Rev. B {\bf 105}, L220202 (2022).
\bibitem{r31}
L. Fleishman and D. C. Licciardello, J. Phys. C: Solid State Phys. {\bf 10}, L125 (1977).
\bibitem{r32}
C.M. Soukoulis and E. N. Economou,  Phys. Rev. B {\bf 24}, 5698 (1981).
\bibitem{r33}
A. Krishna and R.N. Bhatt, Phys. Rev. B {\bf 101}, 224203 (2020).


\bibitem{Referee1}
{\color{black}D.T. Nguyen, T. A. Nguyen, R. Khrapko, D.A. Nolan, and N.F. Borrelli, Sci. Rep. {\bf 10}, 7156 (2020). 
}

\bibitem{r33b}
S.E. Venegas-Andraca, Quantum Inf. Process. {\bf 11}, 1015 (2012).
\bibitem{r34}
 A. Schreiber, K. N. Cassemiro, V. Potocek, A. Gabris, P. J. Mosley, E. Andersson, I. Jex, and
C. Silberhorn, Phys. Rev. Lett. {\bf 104}, 050502 (2010).
\bibitem{r35}
K. Wang, X. Qiu, L. Xiao, X. Zhan, Z. Bian, W. Yi, and P. Xue, Phys. Rev. Lett. {\bf 122}, 020501 (2019).
\bibitem{r36}
A. Regensburger, C. Bersch, B. Hinrichs, G. Onishchukov, A. Schreiber, C. Silberhorn, and U. Peschel, Phys. Rev. Lett. {\bf 107}, 233902 (2011).
\bibitem{r37}
M. Wimmer, A. Regensburger, M.-A. Miri, C. Bersch, D.N. Christodoulides, and U. Peschel,  Nat. Commun. {\bf 6}, 7782 (2015).
\bibitem{r37b}
S. Wang, C. Qin, W. Liu, B. Wang, F. Zhou, H. Ye, L. Zhao, J. Dong, X. Zhang, S. Longhi, and P. Lu, Nature Commun. {\bf 13}, 7653 (2022).
\bibitem{r38}
F. Cardano, F. Massa, H. Qassim, E. Karimi, S. Slussarenko, D. Paparo, C. de Lisio, F. Sciarrino, E. Santamato, R.W. Boyd, and L. Marrucci,
Science Adv. {\bf 1}, e1500087 (2015).
\bibitem{r39}
F. Cardano, A. D'Errico, A. Dauphin, M. Maffei, B. Piccirillo, C. de
Lisio, G. De Filippis, V. Cataudella, E. Santamato, L. Marrucci, M.
Lewenstein, and P. Massignan,  Nature Commun. {\bf 8}, 15516 (2017).
\bibitem{r40}
A. D' Errico, R. Barboza, R. Tudor, A. Dauphin, P. Massignan, L. Marrucci, and F. Cardano, 
APL Photonics {\bf 6}, 020802 (2021).
\bibitem{Referee2}
{\color{black} 
N. Lo Gullo, C.V. Ambarish, Th. Busch, L. Dell'Anna, and C.M. Chandrashekar, Phys. Rev. E 96, 012111 (2017).}
\bibitem{r41}
A. Ahlbrecht, V.B. Scholz, and A.H. Werner,  J.  Math. Phys. {\bf 52}, 102201 (2011).




\end{thebibliography}
\end{document}